\documentstyle[graphicx,twocolumn,aps]{revtex}

\begin{document}

\twocolumn[\hsize\textwidth\columnwidth\hsize\csname@twocolumnfalse\endcsname

\draft

\title{Field distribution and flux-line depinning in the mixed-state of MgB$_{2}$
probed by Conduction Electron Spin Resonance }
\author{R.R. Urbano,$^{1}$ P.G. Pagliuso,$^{1}$ C. Rettori,$^{1}$ Y. Kopelevich,$%
^{1} $ N.O. Moreno,$^{2}$ and J.L. Sarrao$^{2}$}
\address{$^{1}$Instituto de F\'{i}sica \ ''Gleb Wataghin '' UNICAMP, 13083-970 \\
Campinas, SP, Brazil.\\
$^{2}$Los Alamos National Laboratory, Los Alamos, New Mexico 87545, U.S.A.}
\date{\today}
\maketitle

\begin{abstract}
We report the first observation of the internal magnetic-field distribution
and flux-line $depinning$ in the vortex-state of a type-II superconductor
probed by conduction electrons spin resonance (CESR) technique. The CESR
measurements were performed in the recently discovered MgB$_{2}$ type-II
superconductor compound with transition temperature $T_{c}\simeq 39$ K using
microwave sources at $4.1$ (S-band) and $9.5$ GHz (X-band) corresponding to
resonance fields of $H_{r}\simeq 1455$ and $3390$ Oe for $g\simeq 2.00$ in
the normal state, respectively. From the\ distortion of the CESR line in the
superconducting state,\ the field distribution function, $n(H)$, in the
vortex-state was inferred, and from the broadening\ of the line a direct
estimate of the standard deviation, $\sigma \sim $ $14$ Oe, was obtained at $%
\approx $ $28$ K and $\approx $ $7$ K for S-band and X-band, respectively.
Furthermore, our experiments allowed the determination of the flux-line
lattice $depinning$ temperature for both employed microwave frequencies.
\end{abstract}


\vskip 2pc]

\narrowtext

In the early 70's almost simultaneously three groups reported the
observation of electron spin resonance (ESR)\ of localized magnetic
impurities in the mixed-state of type-II intermetallic superconductors.\cite
{Rettori,Engel,Alekseevski} The effects caused by the superconducting state\
on the resonance lineshape, field for resonance ($g$-value), and resonance
linewidth were later discussed in detail by Davidov $et$ $al$. \cite{Davidov}
Concurrently, Orbach \cite{Orbach} showed that the gross observed features
could be explained in terms of the internal magnetic-field distribution in
the Abrikosov vortex-lattice. \cite{Abrikosov} Following the Lasher's
calculations using the Ginzburg-Landau equations\cite{Lasher} and the
analysis of the nuclear magnetic resonance (NMR) data in Vanadium given by\
Fite and Redfield, \cite{Fite,Redfield} Orbach was able to simulate most of
the observed features based on the Abrikosov vortex-lattice internal
magnetic-field distribution of a type-II superconductor. Although\ CESR
experiments in normal metals were discovered in the early 50's, \cite
{Griswold} only in the 80's Vier and Schultz reported the first observation
of CESR in the superconducting mixed-state of the Nb type-II superconductor.
\cite{Vier} However, due to the low $H_{c2}$ [$H_{c2}(T=0)\approx 4$ kOe],
strong decrease of the CESR intensity, narrowing effects, relaxation
phenomena, and weak pinning in the superconducting state, \cite
{Maki,Tsallis,Yafet} these authors have not observed the effects of the
vortex-lattice field distribution in their CESR experiments. Nemes $et$ $al.$
\cite{Nemes} reported the observation of the CESR in the superconducting
state of K$_{3}$C$_{60}$ [$T_{c}(H=0)\approx 19$ K, $H_{c2}(T=0)\approx 25$
T]. Nonetheless, the observed $T$-dependence of the CESR linewidth below $%
T_{c}$ did not allow them to distinguish between the contribution of the
field distribution in the vortex-lattice from that of the relaxation
processes. \cite{Maki,Tsallis,Yafet} More recently, Simon $et$ $al.$ \cite
{Simon} reported CESR in MgB$_{2}$. However, they have observed the
vortex-lattice field distribution either, probably, due to the high
field/frequency used in their experiments.

In this work we report, the first direct and unambiguous observation of the
internal magnetic-field distribution and flux-line $depinning$ in a type-II
superconductor probed by conduction electrons ($ce$). The standard deviation
of the field distribution has been quantitatively estimated from the
experiment.

The recent discovery of superconductivity in the binary compound MgB$_{2}$
at $T_{c}$ $\simeq 40$ K \cite{Nagamutsu} and its high upper critical field $%
20$ T $\lesssim H_{c2}^{\parallel ,\perp c}$ $\lesssim $ $30$ T \cite{Oscar}
have attracted much interest and stimulated us to investigate the
mixed-state of this type-II superconductor by means of CESR. To probe the
vortex-lattice internal magnetic-field distribution by $ce$ we have choosen
our two lowest available microwave sources to perform the CESR experiments.
The S-band ($\nu \approx 4.1$ GHz, $H_{r}\approx 1455$ Oe for $g=2.00$ in
the normal state) and X-band ($\nu \approx 9.5$ GHz, $H_{r}\approx 3390$ Oe
for $g=2.00$ in the normal state) is well suited for this purpose, since for
$T\ll T_{c}$ the CESR field, $H_{r}$, will be above $H_{c1}\lesssim 500$\ Oe
and well below the irreversibility field, $H_{r}\ll H_{irr}^{\parallel
,\perp c}<H_{c2}^{\parallel ,\perp c}$ (see below) and, therefore, the $ce$
will be certainly probing the internal magnetic-field distribution in the
vortex-lattice. Besides, at $\nu \approx 9.4$ GHz and $T\gtrsim 40$ K the
lowest estimates for the skin depth $\delta \gtrsim 1$ $\mu $m [$\delta =$ $%
\sqrt{\rho /\pi \mu _{0}\nu }$, and $\rho $($T$) from ref. \cite{Canfield}]
is larger than the average size of our fine powder particles (our MgB$_{2}$
particle size ranged between $0.5$ $\mu $m and $1$ $\mu $m, determined by
optical microscopy). These two constrains improve the CESR signal/noise
ratio and simplify the analysis of the CESR spectra, since a pure absorption
lorentzian line is expected to be observed at all temperatures. \cite{Dyson}

The MgB$_{2}$ polycrystalline sample was prepared in sealed Ta tubes as
described previously.\cite{Bud'ko} X-ray powder diffraction analysis
confirmed single-phase purity and AlB$_{2}$-type structure for our MgB$_{2}$
sample. The zero field cooled (ZFC) and field cooled cooling (FCC)
diamagnetism was measured by $dc$-magnetization in a SQUID MPMS-QD
magnetometer at $10$ Oe and at the S and X-band CESR fields of $1455$ Oe and
$3387$ Oe, respectively. The CESR experiments were carried out in an
ELEXSYS-CW S and X-band Bruker spectrometer using a flexline probehead line
with a dielectric ring/split ring low Q cavity module for S-band and a TE$%
_{102}$ cavity for X-band. The microwave and external fields were always
mutually perpendicular. The microwave power was keept as low as $0.1$-$0.5$
mW to minimize the unpleasant noise induced by the $ac$-microwave and
modulation fields in the superconducting\ mixed-state and, when necessary, $%
4 $ scans were accumulated to improve the signal/noise ratio. A $100$ kHz
field modulation/lock-in signal detection system and a He gas flux $T$%
-controller were used.

\begin{figure}[h]
\centerline{\includegraphics[scale=0.35,angle=270]{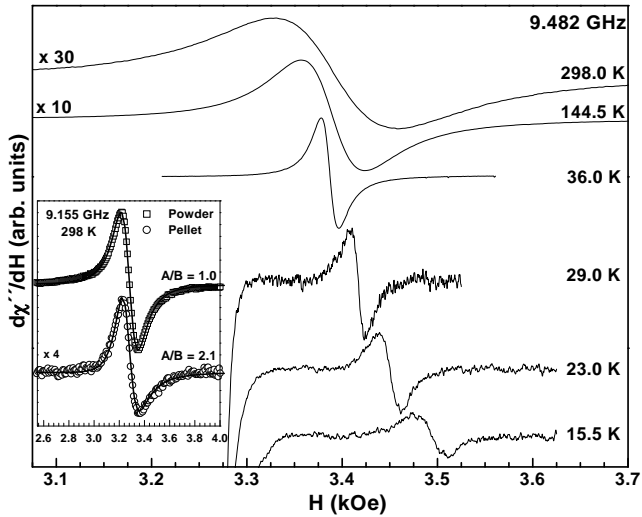}}
\caption{X-band CESR spectra of MgB$_{2}$ at various temperatures
above and below $T_{c}=39$ K. The inset shows, in open simbols,
the X-band CESR spectra at room temperature for a fine powder and
a pellet made out pressing the same powder and the solid lines
are the fittings to lorenztian and dysonian lineshapes.}
\label{Figure 1}
\end{figure}

Figure 1 presents the FCC X-band CESR spectra at few temperatures
above and below the transition temperature $T_{c}(3387$
Oe$)\simeq 36$ K for a fine powder sample of MgB$_{2}$. Above
$T_{c}$\ the lineshape is lorentzian. Calibration of the CESR
intensity against a strong pitch in KCl at room-$T$ leads to
$\approx 3.6$ x$10^{20}$ spin/cm$^{3}$ and a spin susceptibility
of $\chi _{s}\approx 1.4(5)$ x$10^{-5}$ emu/mole. Using a free
electron gas approximation ($\chi =\mu _{B}^{2}N(0)$) we
extracted the density of states (DOS) at the Fermi level,
$N(0)\approx 0.42(5)$ states/eV$\cdot $unit cell, which is of the
order of that obtained from a band structure calculation.
\cite{Kortus} The inset of Fig. 1 shows the X-band CESR spectra
measured at room temperature for a fine powder and for a pellet
made out of the same powder. For the pellet sample we observed a
CESR of dysonian lineshape with A/B $\approx 2.1$ (admixture of
absorption and dispersion of lorentzian line) indicating\ a
sample size larger than or comparable with the skin depth,
$\delta $, \cite{Dyson} and with an intensity smaller than that in
the fine powder.\ These results confirm that the CESR indeed
comes from the bulk of the sample under study. We want to point
out that in all our experiments a single resonance was always
observed. FCC and ZFC X-band CESR in the superconducting
mixed-state at $19.3$\ K\ for the same sample of Figure 1 are
shown in Fig. 2a. It is clear from these measurements that the
shift toward higher fields of the CESR, $\sim 20$ Oe larger for
the ZFC experiment, is caused by the diamagnetic shielding effect
in the superconducting state. Fig. 2b presents the supercoducting
transition measured for this sample by $dc$-magnetic
susceptibility, $\chi _{dc}(T)$, in a ZFC and FCC experiments at
$10$\ Oe, $1455$\ Oe (S-band), and $3387$\ Oe (X-band). The inset
of Fig. 2b shows, for the three\ applied fields, the
onset of superconductivity, $T_{c}(H$), and the irreversibility points, $%
T_{irr}(H)$. For $H=$ $10$ Oe we obtained $T_{c}\simeq T_{irr}=39$ K.

\begin{figure}[tbp]
\centerline{\includegraphics[scale=0.45]{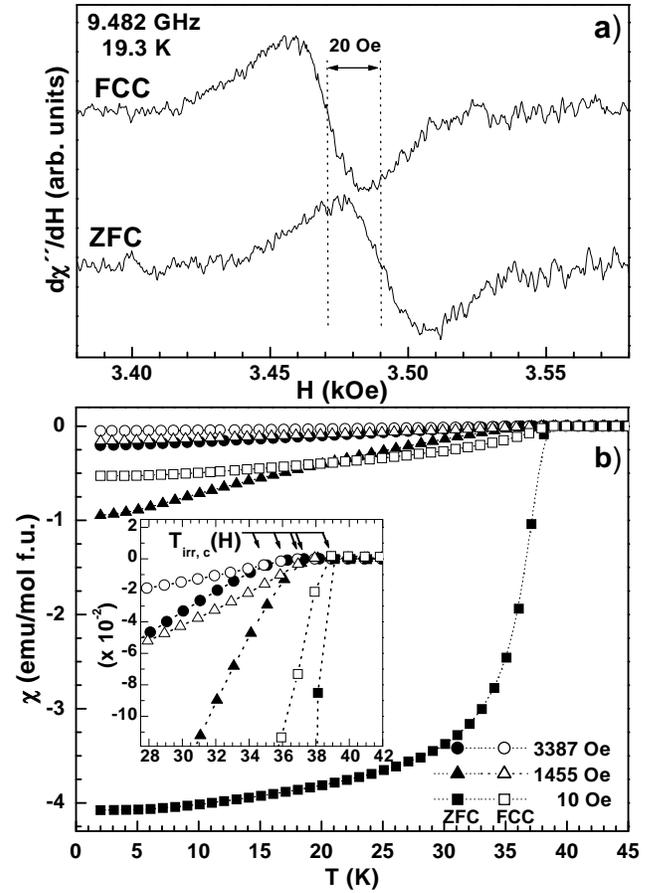}} \caption{a)
FCC and ZFC X-band CESR spectra at $19.3$ K, b) $T$-dependence
of the FCC and ZFC $dc$-magnetic susceptibility of MgB$_{2}$ for $H=10$ Oe, $%
1455$ Oe and $3387$ Oe.}
\end{figure}

Regarding the normal state properties of MgB$_{2}$, we found, for $T\gtrsim
200$\ K and a Pauli-like $T$-independent magnetic susceptibility $\chi
_{dc}(T)$\ $=2.5$\ x$10^{-5}$\ emu/mole, after taking into account the core
diamagnetism. Comparing this value with that\ obtained from\ the DOS
calculated theoretically, $N(0)\approx 0.65(5)$ states/eV$\cdot $unit cell
and using $\chi =$ $\mu _{B}^{2}N(0)$, one can estimates the Stoner%
\'{}%
s factor $\chi _{dc}/\chi =1/(1-\alpha )$ $\approx 1.19$ ($\alpha \approx
0.16(6)$) due to the electron-electron exchange interactions in MgB$_{2}$.\
Estimates of $\chi $ using the Sommerfeld constant $\gamma =1/3\pi
^{2}k_{B}^{2}N(0)(1+\lambda )\approx 2.6$ mJ/mole K$^{2}$,\cite{Bouquet,Wang}
are not reliable due to\ the large uncertainty found in the literature for
the value of the electron-phonon coupling, $\lambda $.\cite
{Patnaik,Yildirim,Cappelluti,Osborn}

Below we discuss the CESR results obtained in a FCC experiments. The $T$%
-dependence of the relative intensity, $I$($T$)/$I^{n}$($40$\ K), the shift
of the field for resonance relative to that in the normal state, $H_{r}(T)$-$%
H_{r}^{n}(40$\ K$)$, and the linewidth, $\Delta H_{pp}$($T$) of the CESR
measured at X and S-bands, are presented in Figs. 3a, b, and c,
respectively, for the sample of Fig. 1. In the normal state, $T\gtrsim 39$
K, we found that the CESR lineshape is lorentzian at all $T$ and, within the
accuracy of the measurements, $I^{n}(T)$ and $H_{r}^{n}(T)$ are $T$%
-independent indicating that the resonance can be attributed to itinerant $ce
$ ($g=2.003(2)$) as expected for a light metal. \cite{Y.Yafet} Fig. 3c shows
that above $\sim 33$\ K the linewidth, $\Delta H_{pp}$, is frequency
independent suggesting homogeneous CESR. Between $40$ K and $110$ K $\Delta
H_{pp}$ follows roughly the $T$-dependence of the resistivity, $\rho $($T$),
examplified by the data reported by Canfield $et$ $al.$ for a MgB$_{2}$
wire. \cite{Canfield} This result suggests that in this temperature region
the $ce$ spin-lattice relaxation is dominated by phonons via spin-orbit
coupling.\cite{Webb,Y.Yafet} However, above $110$ K a clear departure of $%
\Delta H_{pp}$($T$) from $\rho $($T$) is observed, and above $\sim 250$ K
the linewidth levels off at about $130$ Oe.\ This is an intriguing result
because it shows that while the $ce$ mean-free path is still decreasing at
high-$T$ the spin-flip scattering remains about the same.

In the superconducting state, $T\lesssim 39$ K, the CESR data in
both bands present the following features: $i)$ a strong drop of
the CESR signal/noise ratio becoming almost undetectable at $\sim
4.2$ K ($T/T_{c}\approx 0.1$) in X-band (see Fig. 3a), $ii)$ an
evident resonance shift toward higher fields (see Fig. 3b and
inset), $iii)$ the linewidth, $\Delta H_{pp}^{s}$, does not
change down to $\sim 35$ K and $\sim 26$ K for S and X-band,
respectively, keeping approximately the same value as in the
low-$T$ normal state, $\Delta H_{pp}^{n}\simeq 18$ Oe (see Fig.
3c), and $iv)$ a broadening and distortion of the line (with
larger broadening toward the low field side of the resonance) is
observed below $\sim 35$ K and $\sim 26$ K for S and X-band,
respectively. The drop of $I$($T$)/$I^{n}$($40$\ K) for
$T\lesssim 39$ K is attributed to
the decreasing of normal $ce$ excited across the superconducting gap, $%
\Delta $ ($1.8\div 3$ meV).\cite{Buzea} The fraction of normal $ce$ at $%
T\approx 7$ K, $I(7$ K$)/I^{n}$($40$ K) $\simeq $ $0.7$, is relatively
large, which is a consequence of the mixed-state in type-II superconductors.
The increase of $H_{r}$($T$) for $T\lesssim 35$ K in Fig. 3b is caused by
the partial shielding of the external field by the supercurrents (see also
Fig. 1). Below $\sim 50$ K the linewidth, $\Delta H_{pp}$, is the same in
both bands and remains constant at $\approx 18(2)$ Oe down to $35$ K and $26$
K for S and X-bands, respectively, (see Figs. 3c and 4b).

\begin{figure}[tbp]
\centerline{\includegraphics[scale=0.45]{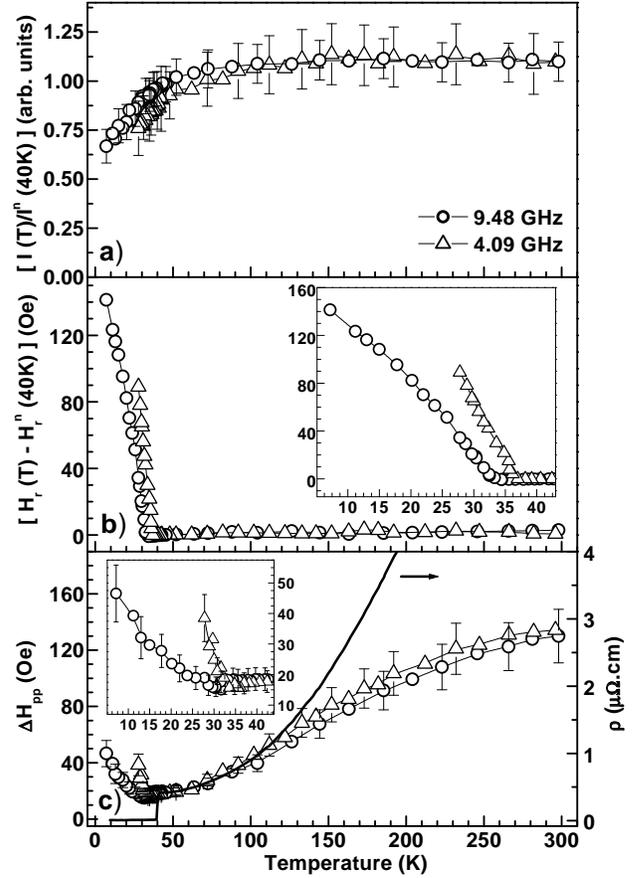}}
\caption{X and S-band $T$-dependence of the: a) relative CESR intensity, $%
I(T)/I^{n}(40$ K$)$, b) shift of the resonance field, $H_{r}(T)$-$%
H_{r}^{n}(40$ K$)$, and c) linewidth, $\Delta H_{pp}$($T$), and
resistivity, $\protect\rho $($T$), from
ref.\protect\cite{Canfield} . }
\end{figure}

Besides, in these $T$-intervals, as in the normal state, the
lineshape remains lorentzian. These results suggest that: $a$)
for\ the S and X-bands, the corresponding temperature intervals
$35\lesssim T\lesssim 37$ K and $26\lesssim T\lesssim 35$ K may
be associated with a vortex-$viscous$ motion regime that, via a
motional narrowing mechanism, may be responsible for the absence
of the inhomogeneous line broadening expected from the
magnetic-field distribution in a vortex-$pinned$ lattice, and
$b$) the $ce$ relaxation in the superconducting state of
MgB$_{2}$ was found to be $T$-independent, except for a tiny
decrease of $\Delta H_{pp}^{s}$\ (smaller than our error bars,
see Fig. 3c).

However, below $\sim 35$ K and $\sim 26$ K for the S and X-band,
respectively, the CESR line clearly broadens and distorts
revealing
the presence of a field distribution that\ we now attribute to a vortex-$%
pinning$ regime. It is worth mention that this broadening cannot be
attributed to a random distribution of the anisotropic upper critical field,
$H_{c2}^{\Vert ,\bot c}$, because at the temperature where the S-band CESR
starts broad the X-band linewidth remains narrow (see Fig. 3c).

The spectra shown in Fig. 4a are the X-band CESR at $\sim 15.5$ K and $%
\sim 23.0$ K. The observed lineshapes present the general features expected
for the CESR absorption derivative of $ce$ probing the internal\
magnetic-field distribution in a vortex-lattice of a type-II superconductor.
Notice that the signal/noise ratio is much smaller than that in the normal
state due to a decreasing number of normal $ce$ in the superconducting state
(see above) and the inevitable noise produced by the interaction between
vortices and the external microwave and modulation $ac$-fields needed for
the detection of the CESR. \cite{Janossy}

\begin{figure}[h]
\centerline{\includegraphics[scale=0.45]{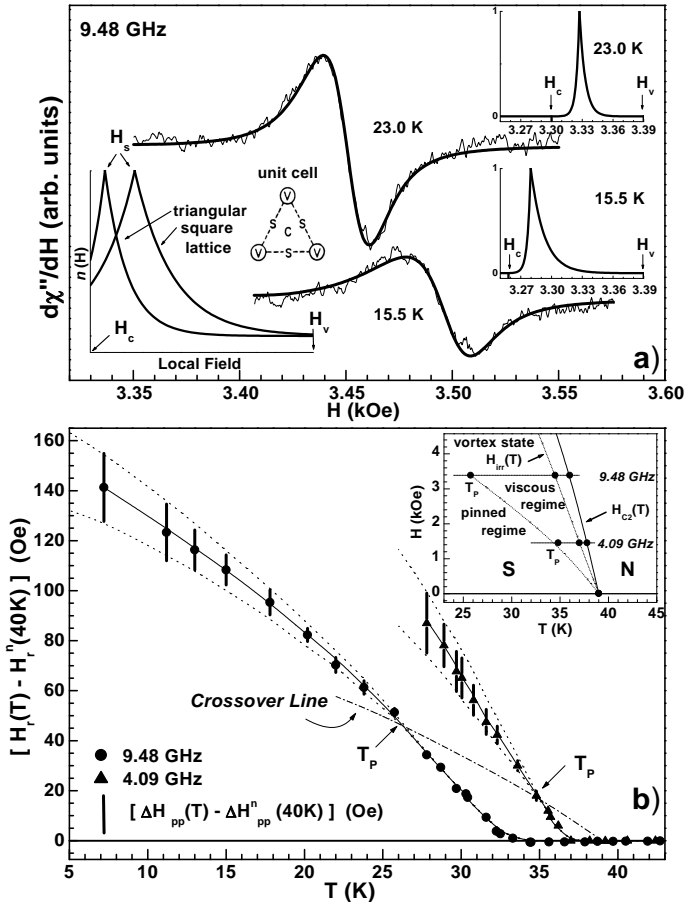}} \caption{ a)
X-band CESR spectra at $15.5$ K and $23.0$ K. Left-bottom
inset: triangular unit cell and Lasher%
\'{}%
s NMR lineshape, $n(H)$, for triangular and square
vortex-lattices. Solid lines in a) are simulations obtained by
the convolution of a lorentzian
absorption line of $\Delta H_{pp}^{n}(40$ K$)=18$ Oe and $H_{r}^{n}(40$ K$%
)=3390$ Oe with a Lasher-like distribution. The best
distributions, $n(H)$,
are shown in the right-side insets. b) $T$-dependence of $H_{r}(T)$-$%
H_{r}^{n}(40$ K$)$ (closed symbols), and $\Delta
H_{pp}(T)$-$\Delta
H_{pp}^{n}(40$ K$)$ (solid bars). Notice that the solid bars are not \ $%
"error$ $bars"$. The inset presents the low field superconducting
state phase-diagram for MgB$_{2}$. The crossover line separating
 vortex-$pinning$ and $viscous$ regimes $T_{p}(H)$ extracted from the CESR data and
irreversibility line obtained from the $dc$-magnetic
susceptibility data(see Fig. 2) are shown.}
\end{figure}

The left-bottom inset of Fig. 4a shows the theoretical \ NMR
absorption lineshape, $n(H),$ calculated by Lasher, using
Ginzburg-Landau equations, \cite{Lasher} due to the internal
magnetic-field distribution in a triangular ($(H_{S}-H_{C})/(H_{V}-H_{C})%
\approx 7\%)$ and square ($(H_{S}-H_{C})/(H_{V}-H_{C})\approx 20\%)$
Abrikosov vortex-lattice, where the points in space correspond to the
maximum, $H_{V}$, minimum, $H_{C}$, and saddle, $H_{S}$ (majority lattice
points) fields in the unit cell of a triangular lattice. For simplicity we
have assumed an Abrikosov vortex-lattice and calculated the derivative of
the convolution of a NMR Lasher-like absorption lineshape, $n(H)$, with an
absorption lorentzian\ lineshape to simulate the CESR absorption derivative
in the superconducting state. In this simulation we have considered: $i)$
for the maximum field, $H_{V}$, the value of the resonance field in the
normal state, $H_{r}^{n}\approx 3390$ Oe, $ii)$ a linewidth, $\Delta
H_{pp}^{s}\approx \Delta H_{pp}^{n}(40$ K) $\approx 18(2)$ Oe (no relaxation
contributions were contemplated \cite{Maki,Tsallis,Yafet}), and $iii)$ two
adjustable parameters, $H_{S}$ and $H_{C}$ ($<H_{S}<H_{V}$) (notice that
CESR is an experiment at fixed frequency, therefore, the diamagnetic shift
of the line will be always toward higer fields). The results of these
convolutions are given in Fig. 4a by the solid lines on the observed
spectra. The best simulations for these two spectra were obtained using the
distribution function, $n(H)$, shown in the right-side insets of Figure 4a.
The reasonable agreement obtained between the data and the simulation
indicates that the broadening and distortion of the CESR lines may be
accounted for by the magnetic-field distribution in the vortex-lattice state
and that relaxation process is not present in the superconducting state of
MgB$_{2}$. According to the results presented in the insets of Fig. 4a the
field distribution deviates from that expected for ideal eitheir triangular
or square vortex-lattice. We attribute this fact to the vortex-lattice
distortions due to the presence of relatively strong vortex pinning effect
in our sample.

Figure 4b presents a summary of the most relevant CESR data in the
superconducting state of MgB$_{2}$. The closed symbols give the shift of the
resonance field in the superconducting state, $H_{r}^{s}(T)$-$H_{r}^{n}(40$ K%
$)$, and the solid bars represent the $"extra"$ broadening of the linewidth
in the superconducting state\ relative to that in the normal state, $\Delta
H_{pp}^{s}(T)$-$\Delta H_{pp}^{n}(40$ K$)$ (notice that these solid bars are
not $"error$ $bars"$). This $"extra"$ broadening may actually be directly
associated to the standard deviation, $2\sigma $, of the field distribution.
The temperature where the resonance field, $H_{r}^{s}(T)$, departures from $%
H_{r}^{n}(40$ K$)$ agrees, within \ $\sim 1$\ K, with the critical
temperature, $T_{c}(H)$, obtained from the onset of superconductivity in
Fig. 2b. The temperature where the $\Delta H_{pp}^{s}(T)$ exceeds that of
the normal state $\Delta H_{pp}^{n}(40$ K$)$, defines the vortex-$pinning$
temperature, $T_{p}(H)$. This temperature separates $pinning$ and $viscous$
vortex motion regime. The microwave field, $H_{1}$, induces a screening
current which exerts a force on flux-lines $j\Phi _{0}$ ($j$ is the current
density) per unit length tilting the flux-lines in the direction of $H_{1}$ $%
(H_{1}\perp H_{r})$. This force is balanced by $pinning$ and $viscous$
forces $\alpha _{p}x+\eta v=j\Phi _{0}$, \cite{RPP}\ where $\alpha _{p}$ is
the $pinning$ constant (Labush parameter), $x$ is the vortex displacement, $%
\eta $ is the $viscous$ drag coefficient (viscosity), and $v$ is the flux
line velocity. So, from the equation of motion one gets the vortex
resistivity $\rho _{v}=(\Phi _{0}H/\eta )/(1+i\omega _{0}/\omega )$\ \cite
{RPP}, where $\omega _{0}=\alpha _{p}/\eta $, the so-called $depinning$
(crossover) frequency, separates the $pinning$ ($\omega \ll \omega _{0}$)
and $viscous$ flux flow ($\omega \gg \omega _{0}$) regimes. We estimate the
in-plane viscosity coefficient $\eta =$ $\Phi _{0}$ $H_{c2}^{\Vert c}/\rho
_{n}\approx 10^{-6}$ Ns/m$^{2}$, taking $H_{c2}^{\Vert c}=2$ T \cite{Simon}
and $\rho _{n}=0.4$ $\mu \Omega $.cm \cite{Canfield}. From low-$T$ ($T=6$ K)
magnetization hysteresis loop $M(H)$ measurements and using the Bean
critical state model applied to a disk-shaped sample, we estimate the
critical current density $j_{c}(T=6$ K, $H=H_{r}=3535$ Oe$)$ $\approx 10^{11}
$A/m$^{2}$ being in a good agreement with previous results \cite{Buzea}.
From the equation for the $pinning$ force $f_{p}=j_{c}\Phi _{0}=\alpha
_{p}\xi $ $(\xi =$ $\sqrt{\Phi _{0}/2\pi H_{c2}^{\Vert c}}$ is the in-plane
coherence length \cite{RPP}$)$ we estimate $\alpha _{p}\approx 1.5$ x$10^{4}$
N/m$^{2}$ , and finally $\omega _{0}\approx 1.5$ x$10^{10}$ rad/s $(\nu
_{0}=\omega _{0}/2\pi \approx 2$ GHz$)$. Because the obtained value of $\nu
_{0}$ is comparable to the measuring frequency, $\nu $ $=9.4$ GHz, and $%
j_{c}(T)$ decreases with $T$, it is reasonable to assume the
vortex $"depinning"$ occurring at $T \geq T_{p}(\nu)$ and motional
narrowing effects may, then, account for the reduction of $\Delta
H_{pp}^{s}(T)$
toward $\Delta H_{pp}^{n}(T)$. If such an interpretation is correct, $%
T_{p}(\nu )$ should be shifted to higher-$T$ by lowering the measuring
frequency. This is actually shown by our experiments at $4.1$\ GHz (see Fig.
4b). On the other hand, for $\nu $ $\gg $ $\nu _{0}$ no crossover to the $%
pinning$ regime is expected and, therefore, no $"extra"$
broadening of the line, $\Delta H_{pp}^{n}(T)$, would be expected
below $T_{c}$. The measurements performed in Ref. \cite{Simon}
for $\nu $ $\geq$ $35$ GHz, which according to our estimation is
well above $\nu _{0}$, revealed at $T=5$\ K a broadening and a
splitting of the line that was attributed to the coexistence of
CESR in the normal and superconducting state. The dashed line in
Fig. 4b is the crossover line which separates $pinning$ and
$viscous$ FLL motion regimes of MgB$_{2}$ obtained from our CESR
experiments. The inset in Fig. 4b presents a low field
phase-diagram for the superconducting state of our MgB$_{2}$
sample extracted from CESR\ and $dc$-magnetization experiments
and summarizes the discussion above.

Finally,\ it is important to mention that the results presented in this
letter and those of Simon $et$ $al.$\cite{Simon} are somehow complementary,
although below $T_{c}$ none of our spectra showed the CESR corresponding to
the normal phase reported in Ref. \cite{Simon}.

In summary, we demonstrate that CESR can be used for direct probing of the
inhomogeneous field dristribution in the mixed state of the MgB$_{2}$. The
standard deviation, $\sigma $, of the field distribution has been inferred
for various temperatures from the $"extra"$ broadening of the linewidth. The
obtained $\sigma $-value of $\sim 14$ Oe at $T$ $\approx $ $28$ K and $%
T\approx $ $7$ K for S-band and X-band, respectively, is of the order of the
values extracted from the analysis of the muon-spin rotation data in high-T$%
_{c}$ cuprates.\cite{Brandt} The small value\ of $\sigma $ ($0.5\div 1\%$ of
the applied fields) is consistent with the relatively large value of the
Ginzburg-Landau parameter $\kappa =\lambda /\xi \sim 10\div 20$ for MgB$_{2}$%
. \cite{Patnaik} Besides, the CESR data has allowed to determine $%
"depinning" $ temperature $T_{p}$($H$, $\nu $) separating vortex-$pinning$
and $viscous$ vortex motion regimes.

The authors are grateful to Prof. O.F. de Lima for critically reading of the
manuscript. This work was supported by FAPESP and CNPq of Brazil.

\end{document}